\documentclass[preprint,amsmath,amssymb,showpacs, aps]{revtex4}

\usepackage[dvips]{graphicx}

\begin{document}

\title{Monte Carlo Simulations of the Critical Properties of a Ziff-Gulari-Barshad model of Catalytic CO Oxidation with Long-range Reactivity}

\author{C.~H. Chan}
\author{P.~A. Rikvold}
\affiliation{Department of Physics, Florida State University, Tallahassee, Florida 32306-4350, USA}

\date{\today }

\begin{abstract}
The Ziff-Gulari-Barshad (ZGB) model, a simplified description of the oxidation of
carbon monoxide (CO) on a catalyst surface, is widely used to study properties of nonequilibrium
phase transitions. In particular, it exhibits a nonequilibrium,
discontinuous transition between a reactive and a CO
poisoned phase. If one allows a nonzero rate of CO desorption ($k$), the line
of phase transitions terminates at a critical point ($k_{c}$). In this
work, instead of restricting the CO and atomic oxygen (O) to react to form
carbon dioxide (CO$_{2}$) only when
they are adsorbed in close proximity, we consider a modified model that
includes an adjustable probability for adsorbed CO and O atoms located far apart on the lattice to react.
We employ large-scale Monte Carlo simulations
for system sizes up to 240$\times$240 lattice sites, using the crossing of fourth-order cumulants to study
the critical properties of this system. We find that the nonequilibrium
critical point changes from the two-dimensional Ising universality class to the mean-field universality class
upon introducing even a weak long-range reactivity mechanism.
This conclusion is supported by measurements of cumulant fixed-point values, cluster percolation probabilities,
correlation-length finite-size scaling properties, and the critical exponent ratio $\beta / \nu$.
The observed behavior is
consistent with that of the \emph{equilibrium} Ising ferromagnet with additional weak long-range
interactions [T. Nakada, P. A. Rikvold, T. Mori, M. Nishino, and S. Miyashita, Phys. Rev. B \textbf{84}, 054433
(2011)]. The large system sizes and the use of fourth-order cumulants also enable determination with improved accuracy
of the critical point of the original ZGB model with CO desorption.
\end{abstract}

\pacs{05.50.+q,64.60.Ht,82.65.+r,82.20.Wt}

\maketitle

\section{introduction}\label{sec:Introduction}
Statistical mechanics has been well developed for equilibrium systems in the sense that one can, in principle,  calculate the partition function and use it to calculate all the equilibrium thermodynamic quantities of a system. On the other hand, the unavailability of the analogy of a partition function for nonequilibrium systems means that nonequilibrium statistical mechanics remains a field in rapid development, attracting researchers to seek its fundamental principles.

In 1986, Ziff, Gulari, and Barshad introduced the ZGB model \cite{PhysRevLett.56.2553} to study the phase transition properties of a particular nonequilibrium process: the formation of carbon dioxide from oxygen and carbon
monoxide at a catalyst surface,
\begin{eqnarray}\label{lamda1}
\rm{CO(g)} + * &&\rightarrow \rm{CO(ads)} \nonumber \\
\rm{O_{2}(g)} +2* &&\rightarrow 2\rm{O(ads)}  \nonumber \\
\rm{CO(ads)} + \rm{O(ads)} &&\rightarrow \rm{CO_{2}(g)} +2*.
\label{eq-1}
\end{eqnarray}
Oxygen ($\rm{O_{2}}$) and carbon monoxide ($\rm{CO}$) gases (g) are supplied to a catalytic surface (Pt). Here, the surface is modeled as a square lattice. When the oxygen molecule ($\rm{O_{2}}$) gets close to the surface, it decomposes into two oxygen atoms ($\rm{O}$). Each O atom and each CO molecule  independently forms a weak bond with an empty lattice site ($*$) to become adsorbed (ads). If a $\rm{CO}$ molecule and an $\rm{O}$ atom are adsorbed at nearest-neighbor lattice sites, they immediately react and form a carbon dioxide molecule ($\rm{CO_{2}}$) that leaves the surface. Each lattice site can either be empty, occupied by one O atom, or occupied by one CO
molecule. The only control parameter in the model is the partial pressure of $\rm{CO}$ in the supplied gas, denoted as $y$. The reaction was simulated by a Dynamic Monte Carlo algorithm, revealing the occurrence of nonequilibrium phase transitions on the catalyst surface. It was found that the steady state of the catalyst surface strongly depends on the partial pressure of $\rm{CO}$ in the feed gas. In this original ZGB model, when the $\rm{CO}$ partial pressure is small, the catalyst surface becomes completely occupied by O atoms in the long-time limit (oxygen-poisoned phase). If the $\rm{CO}$ partial pressure is increased to a value, $y_{1}$, a continuous phase transition occurs, beyond which the catalyst surface is covered by a mixture of $\rm{O}$, $\rm{CO}$, and empty sites (mixed phase).
If one continues to increase the $\textrm{CO}$ partial pressure $y$, a first-order phase transition occurs at $y_{2}$. Beyond this transition, the catalyst surface is completely covered by $\rm{CO}$ in the long-time limit ($\rm{CO}$ poisoned phase).

It has further been noticed in experiments that adsorbed species can desorb from the surface without reacting. The reason for this is that when the temperature is sufficiently high, an adsorbed particle can gain enough energy to break its bond with the catalyst surface. As a consequence, desorption rates increase with temperature. It was also found that the desorption rate of $\rm{CO}$ (denoted as $k$) is much higher than that of $\rm{O}$ atoms \cite{ehsasi:4949}. The desorption rate $k$ of $\rm{CO}$ can be added to the model as a second control parameter \cite{PhysRevA.46.4534,kaukonen:4380,ehsasi:4949,ALBA92}. If a very small value of $k$ is chosen,
there is no qualitative difference in the region of small $\textrm{CO}$ partial pressure. But when $y>y_{2}$, the positive desorption rate reduces the $\rm{CO}$ coverage, producing a nonzero density of vacancies. If the $\rm{CO}$ desorption rate $k$ is increased, a higher $\rm{CO}$ partial pressure, $y_{2}(k)$, is required for the first-order transition to occur. Similar to an equilibrium lattice-gas system, moving along this first-order transition line in the phase diagram eventually leads to a critical point. It has been found that this critical point belongs to the two-dimensional equilibrium Ising universality class \cite{PhysRevE.47.948}.

Soon after the ZGB model was introduced, a number of groups were attracted to study its phase transition properties \cite{PhysRevA.46.4534,PhysRevA.46.4630,PhysRevA.34.4246,PhysRevA.41.3411,meakin:2903,PhysRevE.62.8768,HUA_DaYin:534,Qaisrani:1838,PhysRevLett.77.123}. Some modeled the catalyst surface as a hexagonal lattice instead of a square lattice \cite{meakin:731,PhysRevE.72.066108}. Some studied the effects of oxygen atoms adsorbing at two non-neighboring sites (`hot' dimer adsorption) \cite{doi:10.1142/S0218625X04005846,0305-4470-37-19-002,46436208,0305-4470-27-23-019,Ojeda_and_Buendia}
or as a result of nearest-neighbor repulsive interactions \cite{:/content/aip/journal/jcp/124/15/10.1063/1.2186314,PhysRevLett.84.955,:/content/aip/journal/jcp/111/14/10.1063/1.479949,Liu20091706}. Others considered diffusion of the adsorbed species \cite{LIU04,PhysRevE.65.016121,kaukonen:4380,tammaro:762,:/content/aip/journal/jcp/124/15/10.1063/1.2186314,PhysRevLett.84.955,:/content/aip/journal/jcp/111/14/10.1063/1.479949,Liu20091706}, co-adsorption of the gas molecules (meaning that the gas molecules can react directly with adsorbed species) \cite{buendia:184704}, and the effect of using a periodic $\textrm{CO}$ pressure \cite{Mukherjee20096168,lopez:3890,Buendia2006189,PhysRevE.71.016120}.
Some researchers also studied the effects of impurities present on the catalyst surface \cite{PhysRevE.62.6216,PhysRevE.64.056102} or in the gas phase \cite{PhysRevE.62.8768,HUA_DaYin:534,PhysRevE.85.031143,PhysRevE.88.012132,BUEN14}. Others again
considered the detailed processes happening on the catalyst surface,
building lattice-gas models including energetic effects, with
energy barriers calculated from quantum mechanical DFT calculations and/or comparison with experiments \cite{aip/journal/jcp/114/14/10.1063/1.1343836,Petrova2005162,10.1021/jp071944e,:/content/aip/journal/jcp/124/15/10.1063/1.2186314,aip/journal/jcp/126/4/10.1063/1.2424705,PhysRevB.77.155410,Liu20091706,JCC:JCC22902}.
A recent, comprehensive review of lattice-gas models for CO oxidation on metal (100) surfaces is found in \cite{Liu2013393}, and a review of critical behavior in irreversible reaction systems is found in \cite{LOSC03}.
Other nonequilibrium lattice-gas models with similar phase properties have also been studied \cite{LIU09}.

In equilibrium Ising and lattice-gas
models it is well known that the presence of sufficiently long-range interactions in the Hamiltonian will change the
universality class of the critical point that terminates the line of first-order transitions from the Ising class
to the mean-field class. Given the similarity of the phase diagram of the ZGB model with CO desorption to that of a
liquid-gas system (which also belongs to the Ising universality class), it is natural to ask whether the presence of a
mechanism for long-range reactivity would change the universality class from Ising to mean-field in the same way as
long-range interactions do for the equilibrium  case. A few studies indicate that this is the case.

In one study \cite{0305-4470-27-23-019}, the `hot dimer adsorption' idea was handled by assuming that oxygen molecules
are dissociated and adsorbed as nearest neighbors, but the non-reacted adsorbed oxygen atoms are allowed to undergo a
ballistic flight for up to 20 lattice sites and react with any CO located next to the trajectory. However, the conclusions
of this study regarding universality did not appear very clear.

Much clearer results were obtained in a study by Liu, Pavlenko, and Evans (LPE) \cite{LIU04}, who considered a lattice-gas
reaction-diffusion model similar to the ZGB model with CO desorption, in which adsorbed CO molecules
were allowed to diffuse to
adjacent empty sites at a finite rate, $h$. This leads to an effective diffusion length $\sim h^{1/2}$.
In analogy with earlier studies of equilibrium
Ising systems of linear size $L$ with equal interaction constants of range $\le R$,
which derived a {\it crossover parameter\/} $L/R^2$ for two-dimensional systems
\cite{BIND92,RIKV93,MON93,LUIJ96,LUIJ97,LUIJ97A},
LPE obtained 'effective critical exponents' from finite-size scaling analysis of Monte Carlo simulations. When plotted vs
$L/h$, their results showed good data collapse and a monotonic trend over about two decades of the
crossover parameter
from Ising exponents for $L/h \gg 1$ to mean-field for $L/h \ll 1$. In order to approach the mean-field
limit $L/h \rightarrow 0$ in a computationally manageable way, they resorted to a hybrid model in which the
CO molecules were replaced by a uniform mean field.

In the present paper we approach the problem of determining the universality class of the critical point in a
ZGB model with CO desorption and long-range reactivity along a different path that enables us to
unambiguously extrapolate our results to the limit of infinite-range reactivity and infinite system size.
For this purpose we utilize an analogy with
an approach to the study of phase transitions in Ising-like equilibrium systems with weak long-range interactions,
which has recently been pursued in connection with modeling of spin-crossover materials
with both local and elastic interactions
\cite{PhysRevB.80.064414,PhysRevE.81.011135,PhysRevB.84.054433,NAKA12,PhysRevB.77.014105}.
In this approach,  long-range interactions were added as a perturbation of adjustable magnitude
to an equilibrium Ising system.  Nakada \textit{et al. }\cite{PhysRevB.84.054433,NAKA12} considered a Hamiltonian with both a ferromagnetic nearest-neighbor interaction part (Ising model) and a long-range ferromagnetic interaction part
(the Husimi-Temperley or equivalent-neighbor model in \cite{PhysRevB.84.054433}
and elastic interactions in \cite{NAKA12}).
In both cases they found that, upon the addition of long-range interactions of any nonzero magnitude,
the universality class of the critical point changed abruptly from Ising to mean-field.

Here we  modify the original ZGB model analogously by introducing an adjustable probability that
an O atom and a CO molecule adsorbed far apart on the surface can react to form CO$_2$
and desorb. We use the Random Selection Method of  Dynamic Monte Carlo
\cite{PhysRevE.71.031603} and the crossing of the maximum fourth-order cumulants \cite{PhysRevLett.47.693} to study any resulting changes of the critical properties of the system. In agreement with the results for equilibrium Ising systems \cite{PhysRevB.84.054433,NAKA12},
we find that the universality class of the critical point changes from the Ising class to the mean-field class. In the process, we also obtain an estimate for the critical point of the ZGB model without long-range reactivity which we
believe to be more accurate than, but essentially consistent with, those obtained previously \cite{PhysRevA.46.4534,PhysRevE.47.948,PhysRevE.71.031603}.

The long-range reactivity effect that we introduce here is not intended to model any particular, physical mechanism, but
rather to provide a numerically tractable method to explore the effects of such long-range effects in general.
However, it could be viewed as a simplified version of a rapid diffusion effect \cite{kaukonen:4380,:/content/aip/journal/jcp/124/15/10.1063/1.2186314,PhysRevLett.84.955,:/content/aip/journal/jcp/111/14/10.1063/1.479949,Liu20091706,tammaro:762,PhysRevE.65.016121,LIU04}. Alternatively, practical catalysts are usually made as highly porous materials resembling folded, crumpled surfaces,
so that the
supplied gas encounters a larger surface area per unit volume. In this kind of geometry, it may be possible that an
adsorbed species desorbs and moves to another site, which is far away along the lattice surface, but close in
the three-dimensional embedding space. Our long-range reactivity model could also be viewed as a simplified
model to describe such situations.

 The rest of this paper is organized as follows. In Sec.~\ref{sec:model_and_simulation}, we describe our Monte Carlo scheme and show in detail how we introduce a tunable, long-range reactivity into the ZGB model.
In Sec.~\ref{sec:result}, we show how the phase diagram changes, how we locate the critical point through the
crossing of cumulants
\cite{kaukonen:4380,PhysRevE.71.016120,PhysRevE.71.031603,PhysRevLett.47.693,PhysRevB.84.054433,NAKA12},
and how the universality class of the critical point changes.
In Sec.~\ref{sec:cluster_configuration_and_size}
we provide snapshots for the visualization of the change of
the adsorbate configurations over time, measure the sizes of the largest
cluster in corresponding configurations, obtain the order-parameter
distribution and the cluster percolation probabilities as functions of
CO coverage, and measure correlation lengths and the critical exponent ratio $\beta / \nu$.
Finally, in Sec.~\ref{sec:conclusion}, we summarize our results and state our conclusions.

\section{Model and Simulation}\label{sec:model_and_simulation}

Our study is based on the original ZGB model described in Eq.~(\ref{eq-1}) \cite{PhysRevLett.56.2553}, modified to allow
desorption of $\textrm{CO}$ (i.e., $k>0$) \cite{PhysRevA.46.4534,ALBA92}. In order to include a long-range
reactivity of adjustable strength, the model is modified in the following way. When a
newly adsorbed particle (CO or O) cannot find any
partner particles among its nearest neighbors to react with (O for CO or CO for O), it has a non-zero probability, $a$, to
check a  randomly chosen site anywhere on the lattice. If this site is occupied by a partner particle, the two react to
form CO$_2$ and desorb. In other words, a long-range reaction is considered only after the the possibility of a short-range reaction has been tested and found impossible.
For $a=0$, our model reduces to the standard ZGB model with CO desorption \cite{PhysRevA.46.4534}.
(Although it is known that the unphysical continuous phase transition at $y_{1}$, i.e. the presence
of the oxygen poisoned phase, can be eliminated by considering next-nearest neighbor adsorption instead of nearest-
neighbor adsorption \cite{PhysRevLett.77.123, Ojeda_and_Buendia}, here we stick to the original nearest-neighbor
approach as our focus is the effect of introducing long-range reactivity on the phase transition at $y_{2}$.)
The details of our simulation algorithm are given below.

\subsection{Simulation Algorithm}
\begin{figure}[H]
\includegraphics[width=0.8\textwidth]{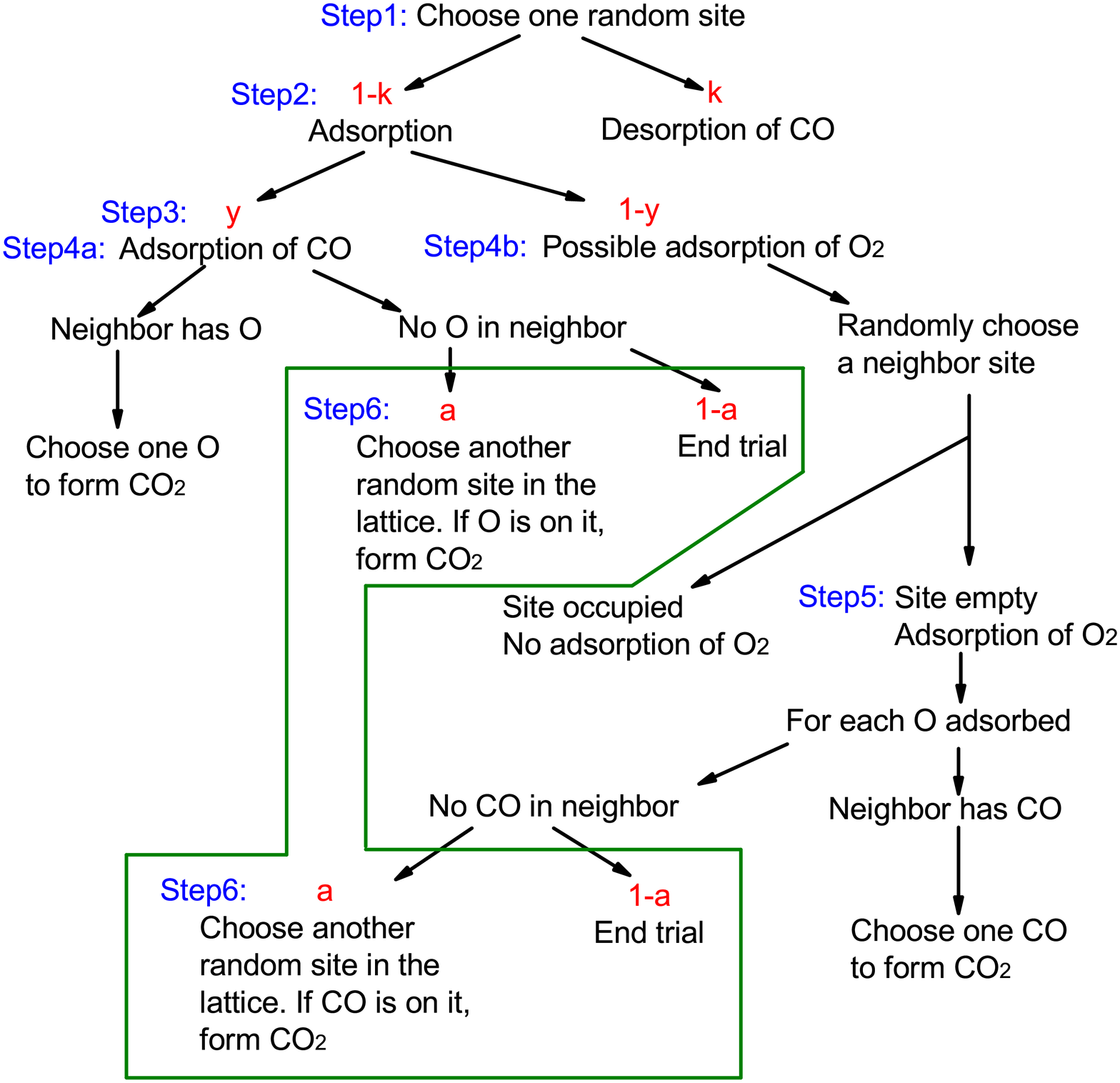}
\caption{(Color online) Flow chart for the reaction process. The algorithm is based on that used in
 \cite{PhysRevE.71.031603} for the ZGB model with CO desorption.
The framed region contains the added long-range reactivity of strength $a$ (Step 6).}
\label{fig_flow_chat_long_range}
\end{figure}

Our algorithm is based on the implementation of
the Random Selection Method of Dynamic Monte Carlo  used in Ref.~\cite{PhysRevE.71.031603}
to simulate the standard ZGB model with CO desorption.
In this method, the whole reaction process is divided into several processes. For each process, there is a separate transition probability. We compare the probability with a random number to decide whether the particular process proceeds or not. A flow chart of the whole process is shown in Fig.~\ref{fig_flow_chat_long_range}. It can be broken down into several steps as follows. The long-range reactivity mechanism is Step~6, which can be reached from Step~4a
if the newly adsorbed particle is a CO molecule, or from Step~5 if the newly adsorbed particle is an O atom.

$\bf{Step~1}$ (choose a site): one lattice site is chosen randomly among the $L\times L$ sites. We do this by drawing a random integer, $r_{1}$, for the $x$-direction and another random integer, $r_{2}$, for the $y$-direction.

$\bf{Step~2}$ (desorption): draw a random real number, $r_{3}\in [0,1]$. If it is smaller than the $\textrm{CO}$ desorption rate ($r_{3}<k\in [0,1]$), and if there is a $\textrm{CO}$ adsorbed at the chosen site, the CO is removed, and this site changes to empty. Then, return to Step~1 for the next trial. On the other hand, if $r_{3} \geq k$ and if this site is empty, go to Step~3. Otherwise, return to Step~1.
The desorption rate $k$ is usually small. (For this work, $0\leq k \leq 0.2$.)  (Note that only the desorption rate of $\textrm{CO}$ is considered, as experiments suggest that it is much greater than the desorption rate of $\textrm{O}$ atoms \cite{ehsasi:4949}.)

$\bf{Step~3}$ (choosing a species to adsorb): draw a random number, $r_{4}\in [0,1]$. If it is smaller than the CO partial pressure ($y\in [0,1]$), then go to Step~4a. Otherwise, go to Step~4b.

$\bf{Step~4a}$ (adsorption of $\textrm{\textrm{CO}}$): if any one of the four nearest-neighbor sites
of this vacant site contains an $\textrm{O}$ atom, the adsorbed $\textrm{CO}$ immediately reacts with
$\textrm{O}$ to form $\rm{CO_{2}}$, which desorbs. If more than one nearest-neighbor site is occupied by
O, draw a random number, $r_{5}$, to choose one of them, and then set both sites to empty
(the original chosen site and this new chosen site). On the other hand, if no O is found at a nearest-neighbor site,
go to Step~6.

$\bf{Step~4b}$ (testing for adsorption of $\rm{O_{2}}$): the $\rm{O_{2}}$ molecule is a dimer. In the ZGB model,
it requires two vacant nearest-neighbor sites for adsorption. To account for the random orientation of the $\rm{O_{2}}$
molecule, we therefore draw a random number, $r_{6}$, to choose one site among the four nearest neighbors of the
originally chosen, vacant site.
If the chosen neighbor is not empty, no adsorption takes place, and we return to Step~1. If the site is
empty, go to Step~5.

$\bf{Step~5}$ (dissociation and adsorption of $\rm{O_{2}}$): the $\rm{O_{2}}$ molecule is dissociated into two $\rm{O}$ atoms and adsorbed. If any one of the nearest neighbors of the first $\textrm{O}$ atom is $\textrm{CO}$, draw a random number, $r_{7}$, to choose one $\textrm{CO}$ among them to react, and then evacuate both sites. If no $\textrm{CO}$ neighbor is found, go to Step~6. Then test the same thing for the second $\textrm{O}$ atom. The trial ends. Return to Step~1.

$\bf{Step~6}$ (long-range reaction): draw a random number, $r_{8}\in [0,1]$. If it is smaller than the long-range reaction probability, $a\in [0,1]$, choose another random site in the lattice. If the two sites contain opposite species (O and CO), they immediately react to form $\rm{CO_{2}}$, which desorbs. The trial ends. Return to Step~1.

In every Monte Carlo step per site (MCSS), we make  $L^{2}$ iterations of the above algorithm with periodic boundary conditions. We choose sufficiently long simulations that the system reaches a steady state, between $5\times 10^{5}$ and $4 \times 10^{8}$ MCSS depending on the parameters, before statistics are taken.

\subsection{Steady state and some properties along the phase boundary}
 A steady state does not mean that the system does not react. Particles can still be adsorbed and react, but certain physical quantities have approached and fluctuate around a steady value. If we consider a region far away from the first-order phase transition region / phase boundary, a steady state means that the coverage of $\textrm{CO}$ ($\theta_{\rm{CO}}$), which is the ratio of lattice sites occupied by $\textrm{CO}$ and is also the order parameter of the system, has reached a steady value. But if we are moving along the first-order phase transition line, due to finite-size effects, the system will jump back and forth between two degenerate stationary states, and thus the $\textrm{CO}$ coverage will repeatedly switch between a high value and a low value. For $k\ll k_{c}$, this switching time can be extremely long. As we increase $k$ towards $k_{c}$, the switching time and the difference between the high and low $\textrm{CO}$ coverages are reduced, while the fluctuations about each stationary level increase. For $k\approx k_{c}$, the fluctuations about the two stationary $\textrm{CO}$ coverages are roughly equal to their separation. This indicates that the system is close to the critical point. Two good quantities to characterize these fluctuations for an $L\times L$ system are
\begin{equation}\label{def_susceptibility}
\chi_{L}=L^{2}( \langle\theta^{2}_{\rm{CO},L}\rangle - \langle\theta_{\rm{CO},L}\rangle^{2}   ),
\end{equation}
(a nonequilibrium analog of equilibrium magnetic susceptibility or fluid compressibility \cite{PhysRevE.71.016120,PhysRevE.71.031603,buendia:184704}),
and the fourth-order reduced cumulant of the order parameter \cite{Landau_simulation_book,PhysRevLett.47.693, PhysRevB.30.1477,PhysRevB.34.1841, PhysRevE.71.031603},
\begin{equation}\label{def_cumulant}
u_{L}=1-\frac{\mu_{4,L}}{3\mu^{2}_{2,L}},
\end{equation}
where
\begin{equation}\label{def_moment}
\mu_{n,L}=\langle(\theta_{\rm{CO},L}-\langle\theta_{\rm{CO},L}\rangle)^{n}\rangle
\end{equation}
is the $n$th central moment of $\theta_{\rm{CO}}$. The `susceptibility' and the cumulant show maxima on the first-order transition line in this system. (Results based on $\chi_{L}$ and $u_{L}$ are consistent. Here we explicitly show only the latter.) Steady state means the system has jumped back and forth many times and has spent the same amount of time at the high level and the low level, so that the susceptibility and the cumulant have been stabilized, but not the coverage.

This switching between the two levels is a finite-size effect. The smaller the system, the easier for the switching to occur and thus the easier it is for the system to stabilize. For an $L\times L$ system, the above simulation process will be repeated $L^{2}$ times. As a larger lattice also makes the physical quantities require more time steps to stabilize, doubling the system size $L$ will make the required running time increase by a factor of more than four. The run times used include $5\times 10^{5}$, $5\times 10^{6}$,  $5 \times 10^{7}$, and $4 \times 10^{8}$ MCSS. The complicated Monte Carlo process and the long time required to stabilize the cumulants make the computation very intensive. More than 600 cores were used for several months to obtain our major results.

\subsection{Initial Conditions}
We chose an initial state with the right half of the lattice sites mainly covered with $\textrm{CO}$ and the left half of the lattice sites mainly covered with $\textrm{O}$. This unstable configuration enabled the system to easily jump very quickly into one of the steady states (around $2000\ \textrm{MCSS}$ for $L=60$).

\section{Cumulants and Phase Diagram}\label{sec:result}

\begin{figure}[h]
\includegraphics[width=1\textwidth]{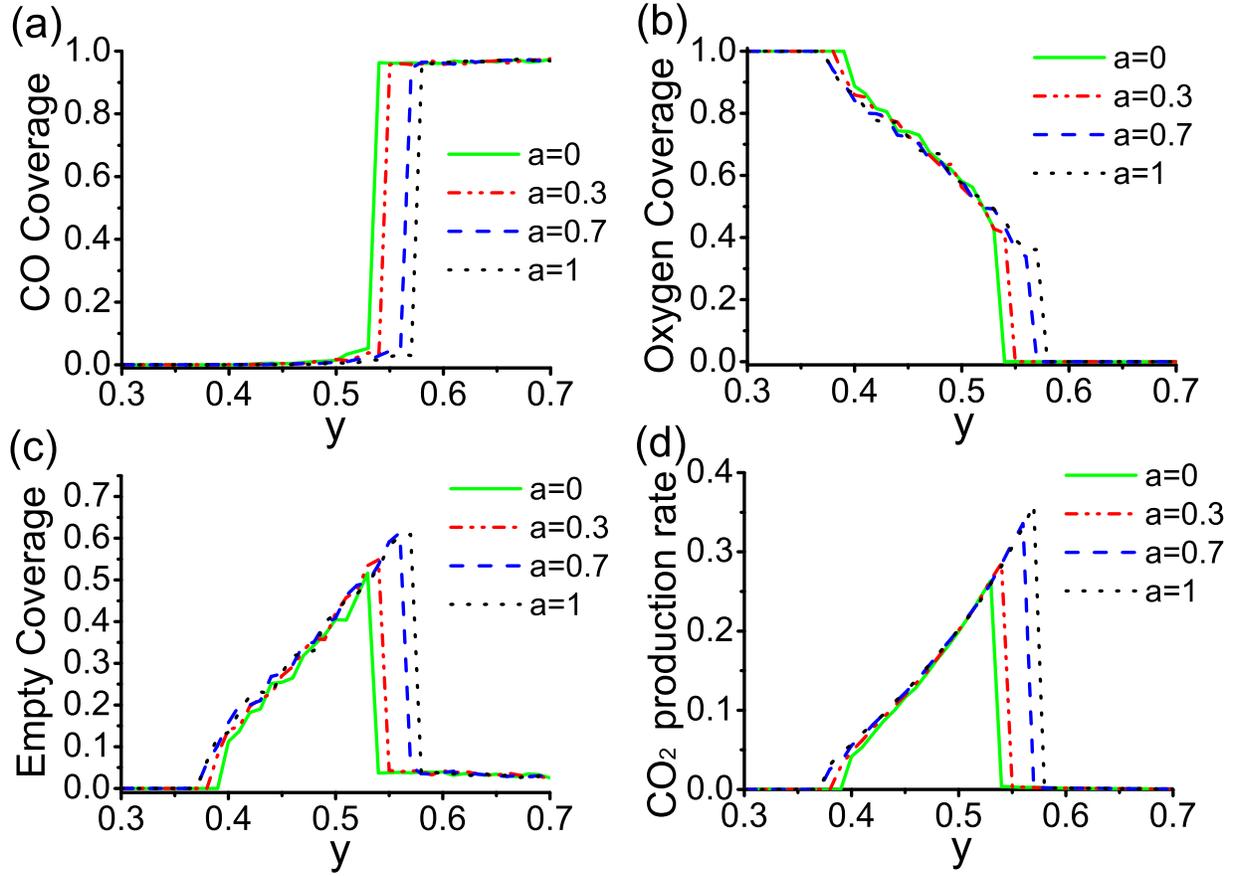}
\caption{(Color online) Coverage ratios of (a) carbon monoxide ($\theta_{\rm{CO}}$), (b) oxygen, (c) empty sites, and (d) $\rm{CO_{2}}$ production rate
on the surface, plotted vs CO pressure in the supplied gas, $y$, using several different values of the long-range reactivity strength $a$. Parameters chosen are CO desorption rate $k=0.02$,
system size $L\times L=60\times60$, and run time $10^{6}$ MCSS. The $\rm{CO_{2}}$ production rate is obtained by averaging the $\rm{CO_{2}}$ produced every 1000 MCSS. The lines for $a=0$ show the results obtained by the original ZGB model with CO desorption, keeping all the other parameters unchanged.
 Data points are taken at intervals $\Delta y =0.01$. }\label{fig_coverage__vs_y_ZGBa_a=all_k=0.02_L=60__IC=2cluster_t=1E6_newboss}
\end{figure}

Figure~\ref{fig_coverage__vs_y_ZGBa_a=all_k=0.02_L=60__IC=2cluster_t=1E6_newboss} shows the coverages and  $\textrm{CO}_{2}$ production rate obtained by our long-range reactivity model for a small desorption value, far below the critical point ($k\ll k_{c}$). We see that increasing the long-range reactivity parameter $a$ from $0$ to $1$ increases the transition point $y_{2}$ by about $7\%$ and the maximum reaction rate by about $36\%$.

\begin{figure}[h]
\includegraphics[width=1\textwidth]{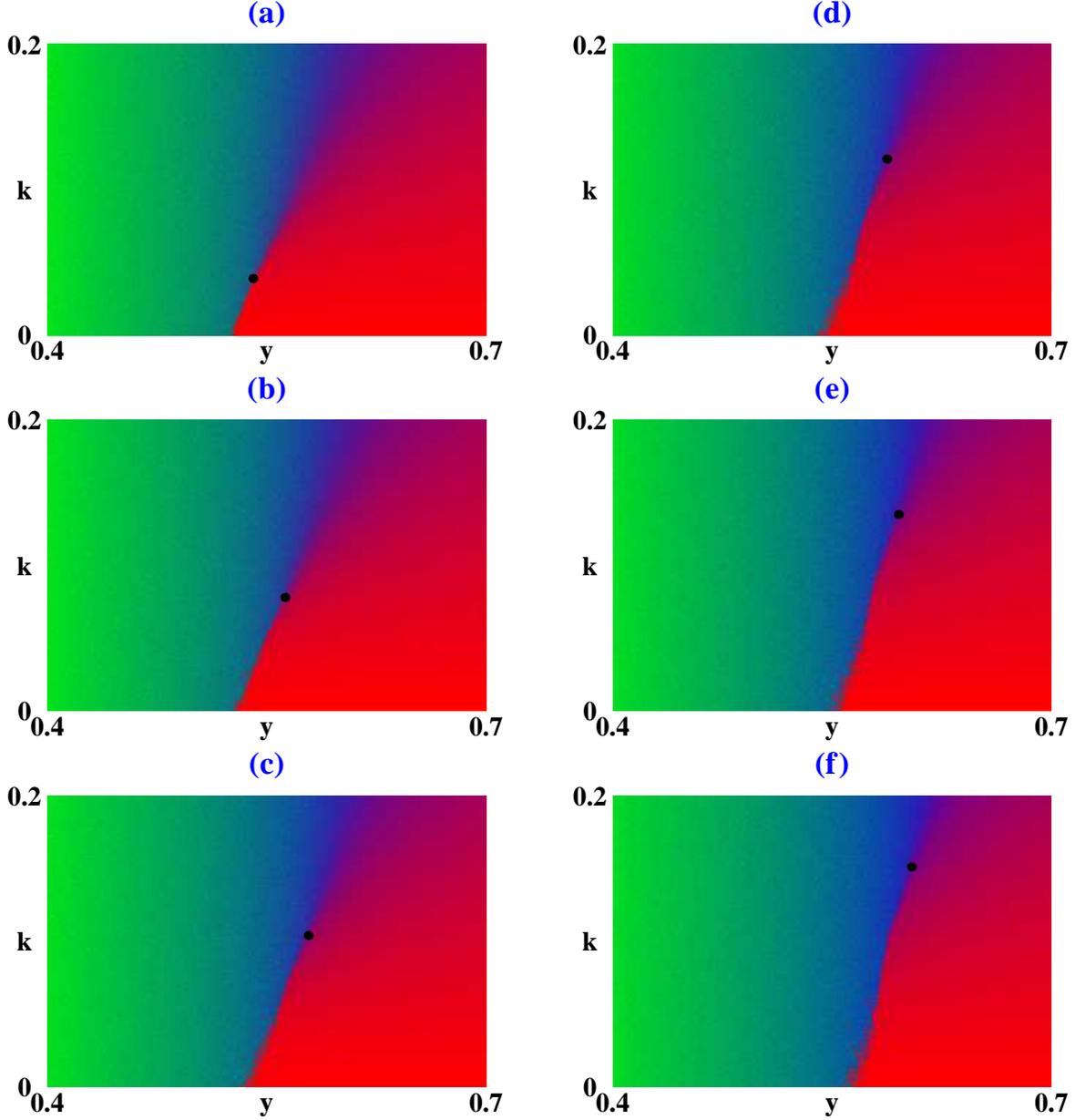}
\caption{(Color online) Phase diagram of the system (a) without ($a=0$) and with ($a>0$) long-range reactivity ( (b) $a=0.1$, (c) $a=0.3$, (d) $a=0.5$, (e) $a=0.7$ and (f) $a=1.0$ ), using $300\times200$ ($k,y$) points, $5\times 10^{4}$ MCSS, and $L=40$. Every point in the phase diagram is composed of 3 different colors, with red representing CO coverage, green representing O coverage, and blue representing empty coverage.
A point in the $(k,y)$ plane with, e.g., CO coverage 0.7, O coverage 0.2, and empty coverage 0.1, is represented by
by a point with color intensities 0.7 red, 0.2 green, and 0.1 blue.
The black dots show the location of the critical point as $L\rightarrow\infty$, obtained from the crossing of cumulants (Fig. \ref{fig_critical_intercept}). In gray scale, the sharp dark line below the black critical point in every diagram is the phase-transition line. The right-hand side of this line is the region where the surface is mainly covered by CO, whereas the left-hand side is the region where it is mainly occupied by oxygen and empty sites with a small density of sites occupied by CO.}
\label{fig_phase_diagram}
\end{figure}

Figure~\ref{fig_phase_diagram} compares the phase diagrams for several values of the long-range reactivity parameter $a\in [0,1]$. The critical point (black dot) moves to a higher desorption rate $k$ and higher $\rm{CO}$ partial pressure $y$ as the long-range reactivity parameter $a$ is increased. Below the critical point ($k<k_{c}$), hysteresis \cite{PhysRevE.66.066103} is found across the first-order phase-transition line.

The first-order phase transition line and the critical point both lie at the value of $y$ that shows a maximum in the cumulants, as shown in Fig.~\ref{fig_a=1_cumulant}(b), \ref{fig_a=1_cumulant}(c), and \ref{fig_a=1_cumulant}(d) for the case with long-range reactivity, and in Fig.~\ref{fig_a=0_cumulant}(b), \ref{fig_a=0_cumulant}(c), and \ref{fig_a=0_cumulant}(d) for the case without long-range reactivity.

\subsection{$a>0$}
\begin{figure}[h]
\includegraphics[width=0.9\textwidth]{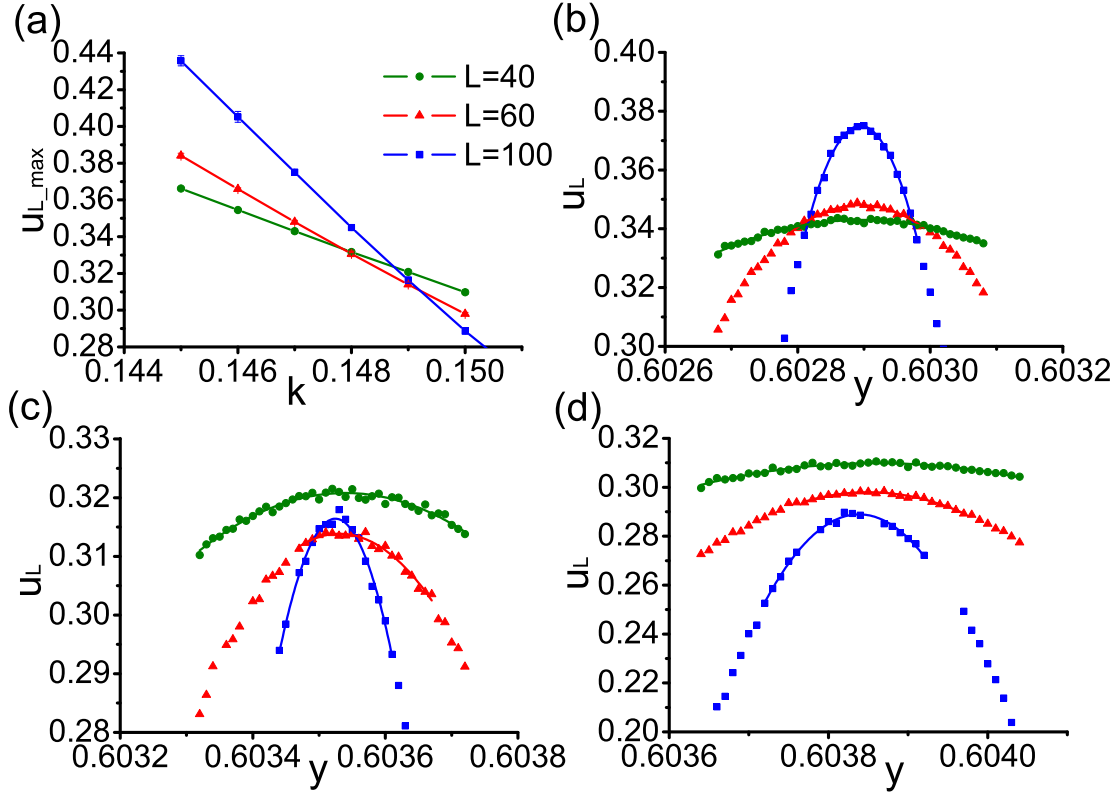}
\caption{(Color online) The search for the critical point ($k_{c,L},y_{c,L}$) for different system sizes
$L$, through the crossing of the maximum of the cumulants $u_{L \rm{max}}$, for long-range reactivity parameter $a=1.0$, using $5\times 10^{7}$ MCSS, $\Delta y=10^{-5}$, with (a) $\Delta k=10^{-3}$, (b) $k=0.147<k_{c,L}$, (c) $k=0.149\approx k_{c,L}$, and (d) $k=0.150>k_{c,L}$. The actual critical point for the infinite-size lattice is found by extrapolation to $1/L=0$ to be $k_{c}=k_{c,\infty}=0.15061\pm 0.00009$, $y_{c}=y_{c,\infty}=0.60402 \pm 0.00003$, as shown in Fig.~\ref{fig_critical_intercept}. (As $k_{c,L}$ increases with $L$, for the system sizes shown here, $L=40,60$, and 100, we get $0.147<k_{c,L}<0.150$.) Error bars for the data points in (a) are comparable to the size of the plotting symbols. They were estimated from the differences between the fitting curve and a narrow range of data points near the cumulant maximum.
 }\label{fig_a=1_cumulant}
\end{figure}

\begin{figure}[h]
\includegraphics[width=0.7\textwidth]{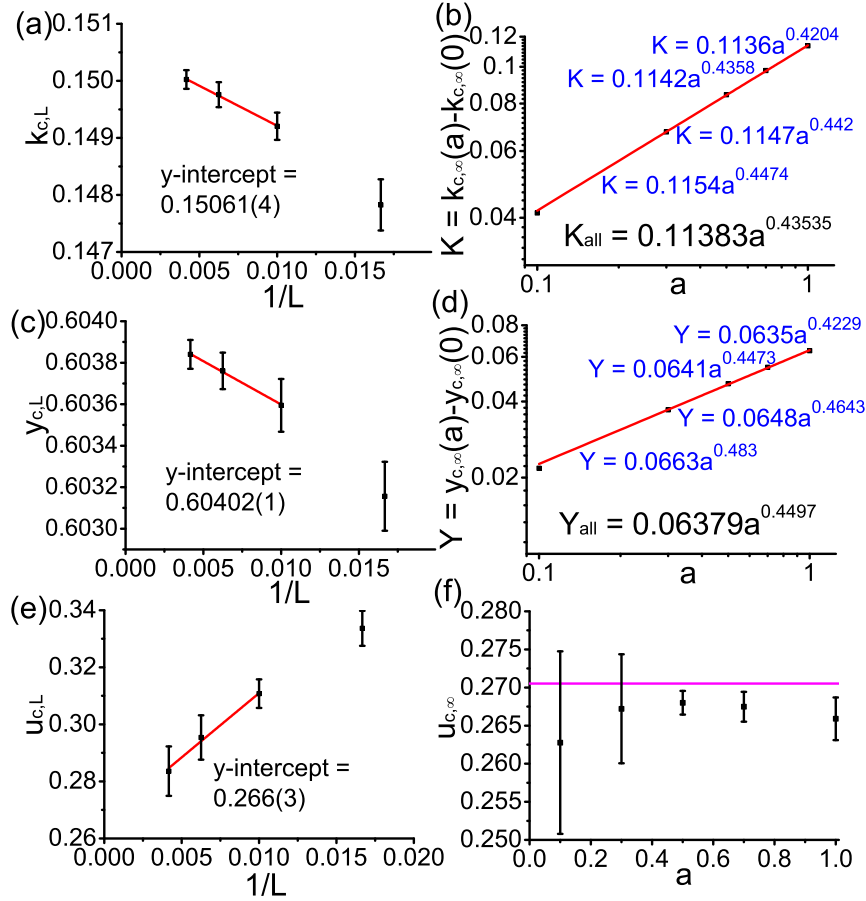}
\caption{(Color online) The search for the critical points ($k_{c},y_{c}$) and the corresponding cumulants $u_{c,L}$ as $L\rightarrow \infty$, for the cases of non-zero long-range reactivity. Lattice sizes $L=40,60,100,160$, and $240$ are used for long-range reactivity strength $a=0.1$, 0.3, 0.5, 0.7, and 1.0 with $5\times 10^{7}$ MCSS,
$\Delta y=10^{-5}$, and $\Delta k=10^{-3}$. Parts (a), (c), and (e) show the case for $a=1$. The crossing point between $L=40$ and $L=60$ in Fig. \ref{fig_a=1_cumulant} was recorded as the $k_{c,L}$, $y_{c,L}$, $u_{c,L}$ for $L=60$.
Similarly, data for $L=100$ is the crossing point between $L=60$ and $L=100$. The $y$-intercepts of the 3 graphs are the critical point ($k_{c},y_{c}$)=($k_{c,\infty},y_{c,\infty}$) and the corresponding cumulant $u_{c,\infty}$ for $L\rightarrow \infty$, and these $y$-intercepts for different values of $a$ were used to obtain the data in parts (b), (d), and (f). The values of $k_{c,L}(0)$ and $y_{c,L}(0)$ used were obtained in Sec.~\ref{sec:a=0}.
 Note that too small system sizes sometimes can deviate greatly from the trend for $L\rightarrow \infty$, so here we did not use the crossing point between $L=40$ and $L=60$ to find $k_{c,\infty}$ and $y_{c,\infty}$. In (b) and (d),
 the equations next to the trendlines are obtained by using every two successive data points, whereas the equations $K_{\rm{all}}$ and $Y_{\rm{all}}$ are obtained by using
  all the data points. The horizontal line in (f) is the exact value ($2.7052...$) of the cumulant for the mean-field universality class \cite{PhysRevB.84.054433}. }\label{fig_critical_intercept}
\end{figure}

\begin{figure}[h]
\includegraphics[width=0.8\textwidth]{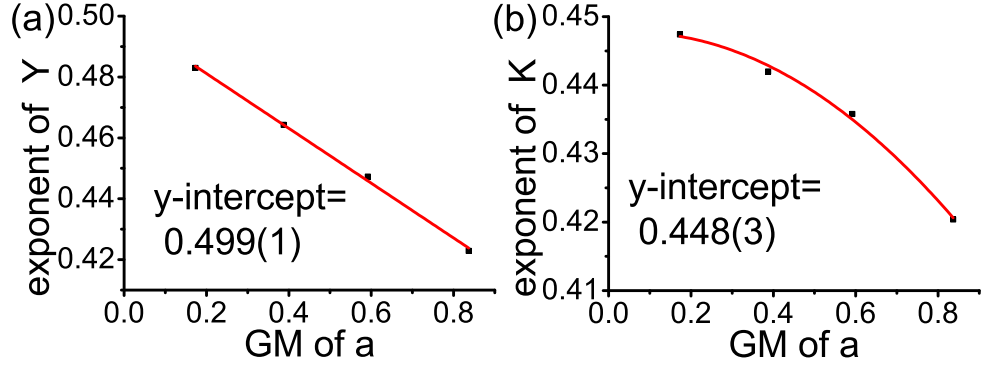}
\caption{(Color online) The search for the exponents in Fig.~\ref{fig_critical_intercept}(b) and \ref{fig_critical_intercept}(d) when $a$ is close to $0$. The $x$-axis is the geometric mean (GM) of the values of $a$ used in Fig.~\ref{fig_critical_intercept}(b) and \ref{fig_critical_intercept}(d). The $y$-intercepts are the resulting exponents, which are quite different from
the behavior $\alpha^{4/7}$ found in the equilibrium Ising model with weak long-range interaction strength $\alpha$ \cite{PhysRevB.84.054433,NAKA12}.}
\label{fig_index_y_k_vs_GM_a}
\end{figure}

We first consider the case with long-range reactivity parameter $a=1$. All the non-zero long-range reactivity cases were found to have similar behavior. Plotting the cumulants against the CO partial pressure, $y$, shows approximately parabolic shapes (Fig.~\ref{fig_a=1_cumulant}(b), \ref{fig_a=1_cumulant}(c), \ref{fig_a=1_cumulant}(d)). The maxima of the cumulants for different system sizes $L$ occur at nearly the same values of $y$.
 For CO desorption rate $k<k_{c}$, the cumulants of different sizes cross each other (Fig.~\ref{fig_a=1_cumulant}(b)), whereas for $k>k_{c}$, the cumulants do not cross (Fig.~\ref{fig_a=1_cumulant}(d)). At $k\approx k_{c}$, the cumulants roughly touch each other (Fig.~\ref{fig_a=1_cumulant}(c)). Due to the fluctuations of the data, we adopted a polynomial fit (2nd-order or 4th-order) to a narrow range of data near the maxima, and used the maxima of the fitting curves as the maximum values of the cumulants. Figure \ref{fig_a=1_cumulant}(a) shows these maximum values of cumulants ($u_{L \rm{max}}$) plotted against the desorption rate $k$ for different system sizes $L$. The line for $L=40$ crosses that for $L=60$ at one point. We picked the two desorption rates just bounding the crossing point, $k_{1},k_{2}$, and used them to form two linear equations that were solved to obtain the crossing point. This crossing point ($k_{c,L},u_{c,L}$) is regarded as the critical desorption rate and its corresponding cumulant found using these two system sizes \cite{PhysRevLett.47.693}, and the index $L$ is taken to be the larger among the two system sizes \cite{Endnote1}. The critical CO partial pressure $y_{c,L}$ found using this two system sizes is obtained from
\begin{equation}\label{y_critical}
y_{c,L}=y_{2}-(k_{2}-k_{c,L})(y_{2}-y_{1})/(k_{2}-k_{1}),
\end{equation}
 where $y_{1}, y_{2}$ are the corresponding values of $y$ for system size $L$, at which the cumulants show maximum values at $k_{1},k_{2}$.
  We recorded the crossing points between every two successive system sizes, and obtained the critical point for $L\rightarrow \infty$ through extrapolation to $1/L=0$ as shown in Fig.~\ref{fig_critical_intercept}(a), \ref{fig_critical_intercept}(c), and \ref{fig_critical_intercept}(e). Figure \ref{fig_critical_intercept}(b) and \ref{fig_critical_intercept}(d) show that $K\equiv k_{c,\infty}(a) - k_{c,\infty}(0)$ and $Y\equiv  y_{c,\infty} - y_{c,\infty}(0) $ both increase in a power-law fashion with the long-range reactivity parameter $a$. ($k_{c,\infty}(0)$ and $y_{c,\infty}(0)$ are obtained in Sec.~\ref{sec:a=0}).
  For the equilibrium Ising model with long-range interaction of strength $\alpha$,
the critical temperature is known to increase as $\alpha^{4/7}$ \cite{PhysRevB.84.054433,NAKA12}.
(4/7 is the Ising critical exponent ratio $\nu / \gamma$.)
The
powers of $a$ observed here are $0.435 \pm 0.004$ and $0.450\pm 0.008$ for $K$ and $Y$ respectively if we use
all the data points in Fig.~\ref{fig_critical_intercept}(b) and \ref{fig_critical_intercept}(d), which are somewhat smaller
than $4/7 \approx 0.571$. We initially suspected this might due to the relatively large minimum value of $a$ used here, so we also found the exponents from the line formed between every two successive data points as shown in Fig.~\ref{fig_critical_intercept}(b) and \ref{fig_critical_intercept}(d). The exponents show a clear increasing trend when $a$ decreases. A polynomial fit was applied to these data as shown in Fig.~\ref{fig_index_y_k_vs_GM_a}. The $y$-intercepts are the exponents we should get when $a$ is non-zero but infinitesimal, which are found to be $0.448 \pm 0.003$ and $0.499 \pm 0.001$ for $k_c$ and $y_c$ respectively, still deviating from $4/7$. Indeed we found that we can obtain $4/7$ only if we use $k_{c,\infty}(0)=0.0515$ and $y_{c,\infty}(0)=0.5472$, which are very far away from the values of $k_{c,\infty}(0)$ and $y_{c,\infty}(0)$ we obtain in Sec.~\ref{sec:a=0} below (see Table~\ref{table_critical_pt}).
One explanation for these results could be that our long-range reactivity parameter $a$ might not be linearly
related to the equilibrium Ising interaction strength $\alpha$.The results for $k_c$ could be reasonably reconciled if
$a \sim \alpha^x$ with $x \approx 1.3$. (Because of the high symmetry of the Ising model, its critical point remains at
zero field for all values of $\alpha$, so
comparing the exponent value for $y_c$ to 4/7 may not be relevant.)


  It is known that in the absence of long-range reactivity, the critical point of the system would correspond to the two-dimensional equilibrium Ising universality class, which has cumulant $u_{c,\infty}\approx0.61$ \cite{0305-4470-26-2-009}. Figure \ref{fig_critical_intercept}(f) shows clearly that for all nonzero values of the long-range reactivity parameter $a$ considered here, the cumulant $u_{c,\infty}\approx0.2675 \pm 0.0009$, consistent with the exact value, $1-\Gamma^{4}(1/4)/24\pi^{2}=0.27052...$, for the mean-field universality class of the equilibrium Ising system with long-range interactions \cite{Brezin1985867, doi:10.1142/S0129183195000265, PhysRevB.77.014105,PhysRevB.84.054433}. Table~\ref{table_critical_pt} summarizes the critical points and the corresponding cumulants obtained for different long-range reactivity strengths $a$. The longest run time we used for the $a=0$ case is $4\times 10^{8}$ MCSS and
for the $a>0$ cases it is $5\times 10^{7}$ MCSS.

\begin{table}
\caption{Critical points and the corresponding cumulants for different values of the long-range reactivity strength $a$. Uncertainty in the last digit given in parenthesis. The asterisks are explained in the Endnote \cite{Endnote1}. }
\centering
\begin{tabular}{llll}
\hline
\hline
$a$ \ \ \  & $k_{c,\infty}$ \ \  & $y_{c,\infty}$\ \  & $u_{c,\infty}$ \\
\hline
0  \ \ \     &   $0.0371(2) $  \ \     &   $0.54052(9) $ \ \   &   $0.624(3) $    \\
0.1 \ \ \     &   $0.0783(4)$  \ \   &   $0.5623(2)$ \ \    &   $0.26(1)$     \\
0.3 \ \ \    &   *$0.1044(1)$ \ \    &   *$0.57758(4)$ \ \    &   $0.267(7)$     \\
0.5 \ \ \     &   $0.1215(2)$ \ \    &   $0.58750(7)$ \ \    &   $0.268(2)$     \\
0.7 \ \ \     &   *$0.1348(1)$  \ \   &   *$0.59513(3)$ \ \    &   $0.267(2)$     \\
1.0 \ \ \     &   *$0.1506(1)$ \ \    &   *$0.60402(3)$ \ \    &   $0.266(3)$     \\
\hline
\end{tabular}\label{table_critical_pt}
\end{table}


\subsection{$a=0$}\label{sec:a=0}
\begin{figure}[h]
\includegraphics[width=0.9\textwidth]{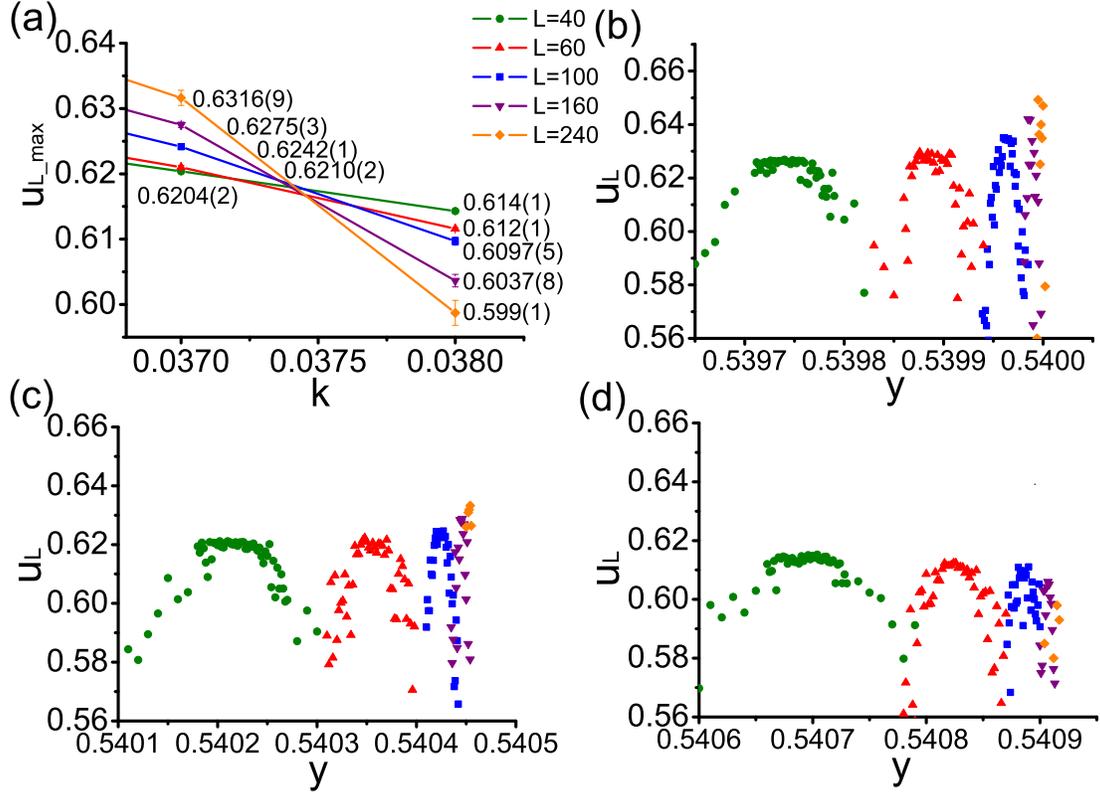}
\caption{(Color online) The search for the critical points ($k_{c},y_{c}$) through the crossing of the maximum of the cumulants $u_{L \rm{max}}$, for the case without long-range reactivity, i.e., $a=0$.
 System sizes $L=40,60,100,160$, and $240$ with $\Delta k=10^{-3}$ were considered.
 $\Delta y=2\times10^{-6}$ was used for $L=40$ and $60$, and $\Delta y=10^{-6}$ was used for $L=100, 160$, and $240$. For $L=40$ and $60$, $5\times 10^{7}$ MCSS were used. For $L=100$, $4\times 10^{8}$ MCSS were used for $k=0.037$, and $5\times 10^{7}$ MCSS for $k=0.038$. For $L=160$ and $240$, $4\times 10^{8}$ MCSS were used. (a) shows the crossing of the maximum of the cumulants $u_{L \rm{max}}$. The numbers shown are the data points and error bars at $k=0.037$ and $k=0.038$. The maximum regions of the cumulants for different system sizes are shown in (b) for $k=0.036<$ crossing point, (c) for $k=0.037\approx$ crossing point, and (d) for $k=0.038>$ crossing point.
Plateaus were found around the maximum regions of the cumulants for all system sizes in (b), (c), and (d).
Error bars in (a) were estimated from the fluctuations of the data points in the plateau regions.
 }\label{fig_a=0_cumulant}
\end{figure}

\begin{figure}[h]
\includegraphics[width=1\textwidth]{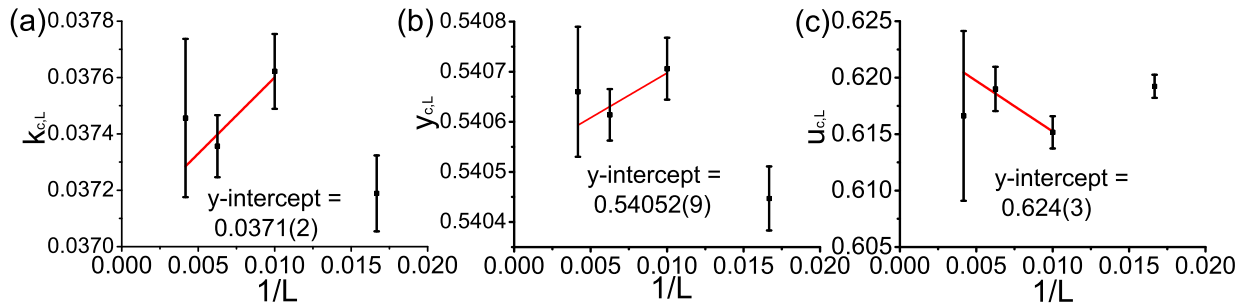}
\caption{(Color online) The search for the critical point ($k_{c},y_{c}$) and the corresponding cumulant $u_{c,\infty}$ for $L\rightarrow \infty$ for the case without long-range reactivity, $a=0$. Lattice sizes $L=40,60,100,160$, and $240$ were considered with $\Delta k$, $\Delta y$ and simulation times chosen as in Fig.~\ref{fig_a=0_cumulant}. Similar to Fig.~\ref{fig_critical_intercept}, the crossing point between $L=40$ and $L=60$ in Fig.~\ref{fig_a=0_cumulant} was recorded as the data point of $L=60$. The $y$-intercepts of these three graphs are the critical point ($k_{c},y_{c}$) and the corresponding $u_{c,L}$ for $L\rightarrow \infty$.
Again, too small systems can deviate greatly from the trend for $L\rightarrow \infty$, so
the crossing point between $L=40$ and $60$ (data at $1/L=0.0167$ in the graphs) was not used to obtain the $y$-intercepts of these graphs. }\label{fig_critical_intercept_a0_3graph}
\end{figure}

Figures \ref{fig_a=0_cumulant} and \ref{fig_critical_intercept_a0_3graph} show graphs corresponding to
Figs.~\ref{fig_a=1_cumulant} and  \ref{fig_critical_intercept}, respectively, for the case without long-range reactivity.
Plateaus were found around the maximum regions of the cumulants for all system sizes as shown in Fig.~\ref{fig_a=0_cumulant}(b), \ref{fig_a=0_cumulant}(c), and \ref{fig_a=0_cumulant}(d). Note that even $L=240$ has a plateau. When the system size increases, the plateau moves to a larger value of $y$, and its width decreases. The data points on the plateau also fluctuate more strongly as $L$ increases. For CO desorption rate $k<k_{c}$, the maximum value of the cumulant increases with increasing $L$ (Fig.~\ref{fig_a=0_cumulant}(b)),  whereas for $k>k_{c}$, the maximum value decreases with increasing $L$ (Fig.~\ref{fig_a=0_cumulant}(d)).
At $k\approx k_{c}$, the maximum cumulant value is approximately independent of $L$ (Fig.~\ref{fig_a=0_cumulant}(c)).
The absence of long-range reactivity ($a=0$) leads to larger critical fluctuations that make the system much more difficult to stabilize. The data we obtained in this case were not stabilized as well as those in the long-range reactivity cases.
For the data points shown on the plateaus in Fig.~\ref{fig_a=0_cumulant}(b), \ref{fig_a=0_cumulant}(c), and \ref{fig_a=0_cumulant}(d), the change of the cumulants with time were checked one by one. By looking at the trend of the fluctuating cumulant, we estimated the final stationary value of the cumulant with an error bar for each individual data point (not shown). Then we selected a group of data points near the largest data point, and used the square of the reciprocal of the error as the weight of each data point to find the weighted mean and its standard error. We took these as the maximum value of the cumulant ($u_{L \rm{max}}$) of each curve and its corresponding error bar in Fig.~\ref{fig_a=0_cumulant}(a). The idea in Fig.~\ref{fig_a=0_cumulant}(a) is exactly the same as that in Fig.~\ref{fig_a=1_cumulant}(a). The crossing point between lines for every two successive system sizes $L$ is regarded as the critical point ($k_{c,L},y_{c,L}$) and the corresponding value of $u_{c,L}$ is found using these two system sizes \cite{PhysRevLett.47.693}, and the critical point for $L\rightarrow \infty$ is obtained through extrapolation to $1/L=0$ as shown in Fig.~\ref{fig_critical_intercept_a0_3graph}(a) and \ref{fig_critical_intercept_a0_3graph}(b). $(k_{c},y_{c})=(k_{c,\infty},y_{c,\infty})=(0.0371 \pm 0.0002, 0.54052 \pm 0.00009)$ was finally obtained as the critical point for the case without long-range reactivity. This estimate should be more accurate than previously obtained values \cite{PhysRevA.46.4534,PhysRevE.47.948,PhysRevE.71.031603}, as we used the method of cumulant crossings and the maximum system size was increased to $L=240$. Figure~\ref{fig_critical_intercept_a0_3graph}(c) shows that the maximum value of the cumulant for $L\rightarrow \infty$, is $u_{c,\infty}=0.624\pm 0.003$. Given the numerical difficulties of the simulations of this model for $a=0$, we feel this value is in reasonable agreement with the Ising value of approximately $ 0.61$ \cite{0305-4470-26-2-009}.

In the process of comparing our numerical estimate for $k_c$ at $a=0$ with previous studies, we realized that the
algorithms used in different studies lead to slightly different definitions of the desorption rate \cite{ZIFFpriv}.
While our definition is the same as in \cite{PhysRevE.71.031603}, it is different from the one
used in \cite{PhysRevA.46.4534} and also in \cite{PhysRevE.47.948}. Calling the definition used in
\cite{PhysRevA.46.4534} $P$, the relationship is $P = k/(1-k)$. Consequently, our estimate for $k_c$ corresponds
to $P_c = 0.0385 \pm 0.0002$.  This is close to the approximate  lower bound obtained  in \cite{PhysRevA.46.4534}
from the fractal interface structure, $P_c > 0.039$.

\section{Cluster configurations, cluster-size, and correlation length measurements}\label{sec:cluster_configuration_and_size}

It has previously been demonstrated that the critical configurations are dramatically different in
equilibrium Ising models with
short-range interactions (Ising universality class) and long-range interactions (mean-field universality class).
While the correlation length diverges at the critical point in the former case, it remains finite in the latter
(see, e.g., \cite{PhysRevB.84.054433,NAKA12}).
Visually it is also clear that the Ising critical clusters are larger and more compact than
the mean-field ones (see, e.g., Figs.~5--7 of \cite{PhysRevB.84.054433}).

\begin{figure}[h]
\includegraphics[width=0.5\textwidth]{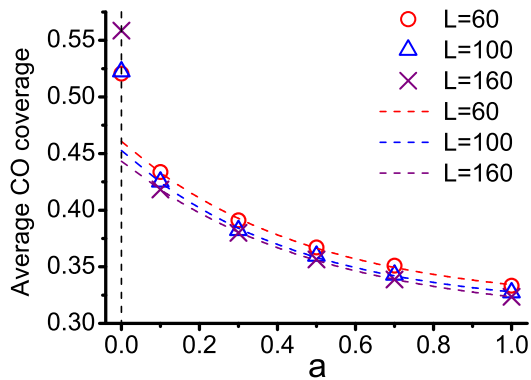}
\caption{ (Color online) Time-averaged, critical CO coverage, shown vs the long-range reactivity strength $a$, for
$L=60$, $100$, and $160$ near their corresponding critical points (at the $60/40$, $100/60$ and $160/100$
cumulant crossings, respectively). $10^{7}$ MCSS were used for $a>0$ and $10^{8}$ MCSS for $a=0$.
The error bars, not shown, are smaller than the symbol size. The dashed curves represent exponential fits to
the $a>0$ data points for different system sizes.
}
\label{fig_critical_CO_coverage}
\end{figure}

We would like to determine whether analogous differences can be observed in the present nonequilibrium system.
However, the high symmetry of  Ising lattice-gas
models ensures that the time-averaged critical coverage is always 1/2, regardless of the strength of the long-range
interactions.
This symmetry does not exist in the model discussed here. Rather,
we find that the critical CO coverage is a decreasing
function of the long-range reactivity strength $a$, as shown in Fig.~\ref{fig_critical_CO_coverage}.
Since cluster properties are
strongly dependent on the coverage, this makes it more difficult to compare critical cluster properties for different
values of $a$.

To solve this problem, we ran simulations of up to $10^8$ MCSS for $a=0$ and $10^7$ MCSS for $a>0$ at their
respective critical points, sampling snapshots every $200$ or $20$ MCSS,
and classified the snapshots according to their CO coverage in
bins of width $0.01$. This enabled us to compare the nonequilibrium
Ising and mean-field critical cluster structures at similar CO coverages. The results are discussed below.

\subsection{Cluster configurations}\label{sec:cluster_configuration}
\begin{figure}[h]
\includegraphics[width=0.5\textwidth]{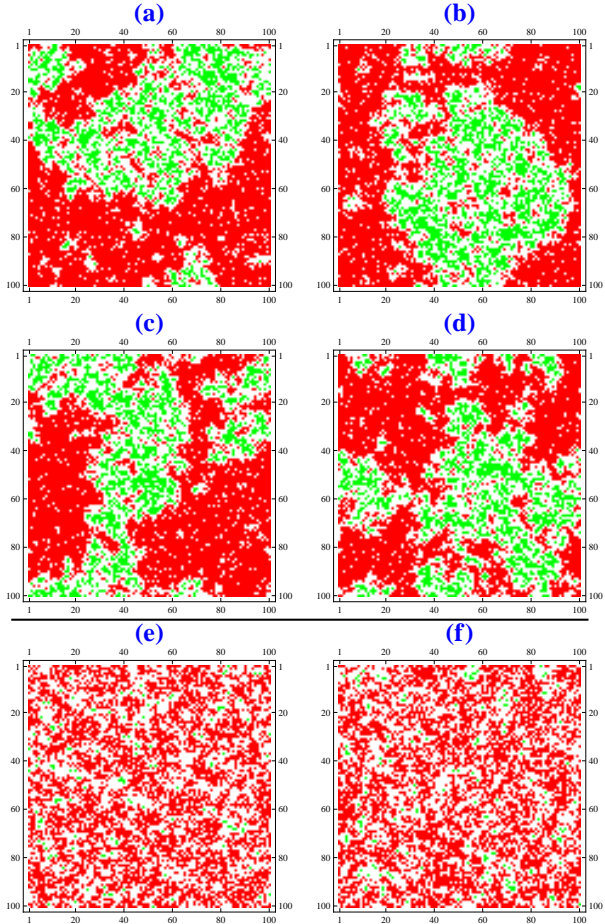}
\caption{(Color online) Snapshots of the adsorbate configurations
for $L=100$ near the corresponding critical point (i.e., at the $100/60$ cumulant crossing)
at different times ($t$ in MCSS) when the CO coverage is near $0.5$ (between $0.48$ and $0.52$).
The upper two rows show the case of $a=0$ ($k=0.0376013,y=0.540699$) at $t=$ (a) $1062400$, (b) $2326000$,
(c) $4541200$ and (d) $5518200$. The bottom row shows the case of $a=1$ ($k=0.149209,y=0.603602$) at $t=$ (e)
$2081640$ and (f) $2081660$. Lattice sites occupied by CO, O, and empty sites are colored as red (dark gray), green (light gray), and white, respectively. All four snapshots for $a=0$ contain a spanning CO cluster.
For $a=1$ in (e) and (f), snapshot (f) was taken just $20$ MCSS after (e).
While the CO coverage of (e) is $0.4848$ and of (f) is $0.4869$, there is a spanning cluster in (e) but not in (f).
Indeed, the radius of gyration of the largest CO cluster in (f) is only $19.2$. This is
essentially impossible to distinguish by visual inspection alone. }
\label{fig_snapshot_critical}
\end{figure}

Figure \ref{fig_snapshot_critical} compares snapshots without and with long-range reactivity
near the critical point for a $100 \times 100$ system at CO coverages close to $0.5$. We see that with the long-range reactivity parameter $a=1$, the clusters are in general smaller or have more empty sites inside big clusters, compared to the case without long-range reactivity, $a=0$. This effect can be easily understood. If a big cluster is formed in the $a=0$ case, the cluster can only change at its boundary, whereas in the $a=1$ case particles in the interior of the cluster can also react with the opposite species outside the cluster to form $\rm{CO_{2}}$ and desorb.
Therefore, in the $a=1$ case an original big cluster will easily be broken up into many small clusters or become a big
cluster with many holes. Moreover, the additional long-range reactivity makes the time required to switch between the high CO state and the low CO state much shorter, as shown in Fig.~\ref{fig_cluster_size_gyration}(a) and Fig.~\ref{fig_cluster_size_gyration}(b). These results are consistent with those obtained by Nakada \textit{et al.} \cite{PhysRevB.84.054433} for the nearest-neighbor Ising ferromagnet and the Ising ferromagnet with weak long-range interactions, respectively.

\subsection{Cluster-size measurements}\label{sec:cluster_size_measurement}
\begin{figure}[h]
\includegraphics[width=0.8\textwidth]{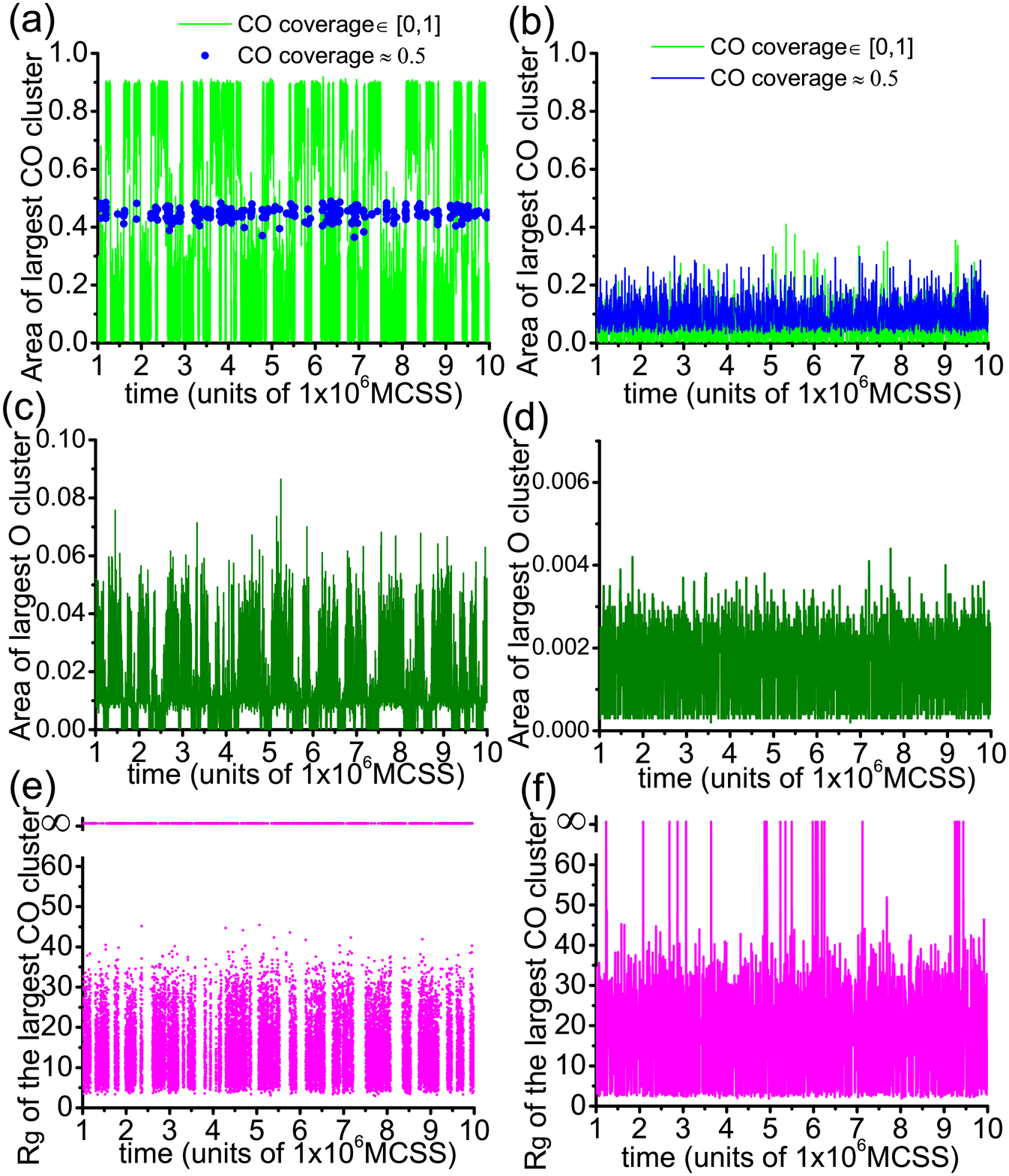}
\caption{(Color online)
Measurement of the areas and the radii of gyration ($R_{g}$) of the largest clusters vs time
for a $100\times 100$ lattice near its critical point (i.e., at the $100/60$ cumulant crossing).
The area is the fraction of the surface occupied by the cluster.
Parts (a), (c), and (e) show the results for $a=0$, whereas (b), (d), and (f) show the results for $a=1$.
Clusters that span the lattice (infinite size under periodic boundary conditions) are recorded as having $R_{g}=\infty$ in (e) and (f).
Data are taken every $200$ MCSS for $a=0$ and every $20$ MCSS for $a=1$, so that $45000$ data points were recorded and shown in (a), (c) and (e), while $450000$ data points were recorded in (b), (d) and (f), only $1/5$ of them are shown for fast display. In (a) and (b), the blue (dark gray) line/dots show the size of the maximum CO cluster only when the system has a CO coverage in the range $0.50 \pm 0.02$, whereas the green (light gray) line shows data with the size of the maximum CO cluster can be in any value.
Among all the samples, $20118$ ($44.71\%$) have a spanning cluster for $a=0$ (a), and only $23$ ($0.051\%$)
have a spanning cluster for
$a=1$ (b). All the spanning clusters are found to be CO spanning clusters, i.e., there are no oxygen spanning clusters.
The time-averaged CO coverage for $a=0$ over $10^{8}$ MCSS is $0.52235$
and for $a=1$ over $10^{7}$ MCSS is $0.32713$.
}\label{fig_cluster_size_gyration}
\end{figure}

A cluster that has infinite size under periodic boundary conditions is called a spanning or percolating cluster
(here defined as one that wraps around the system in one or both directions).
It is interesting to compare the probabilities of finding spanning clusters at comparable CO
coverages in the two cases of $a=0$ (Ising) and $a=1$ (mean-field).
To answer this, we labeled all the CO and O clusters in every configuration using the Hoshen-Kopelman algorithm \cite{PhysRevB.14.3438,Tjipto_thesis}. After the labeling, we measured the sizes of the the largest CO and O clusters vs time as shown in Fig.~\ref{fig_cluster_size_gyration}(a)$-$\ref{fig_cluster_size_gyration}(d). Meanwhile, we measured the radius of gyration of the largest cluster in every configuration as
\begin{equation}\label{radius_of_gyration}
    R_{g} = \sqrt{\frac{1}{2N^2}\sum_{i,j}(\textbf{r}_{i}-\textbf{r}_{j})^{2}},
\end{equation}
where $N$ is the size of the cluster, and $\textbf{r}_{i}$ is the coordinate of a lattice point inside the cluster. Note that due to the periodic boundary conditions, $(x,y)$ and $(x\pm L, y\pm L)$ refer to the same lattice point.
We therefore have to choose the coordinates such that the lattice points are connected through the cluster. To do this, we picked one lattice point inside the cluster and performed a restricted random walk, such that the walker could only walk inside the cluster. Whenever the walker reached a site that had not been visited before, we would assign it a consistent coordinate. Figures~\ref{fig_cluster_size_gyration}(e) and \ref{fig_cluster_size_gyration}(f) show the radii of gyration of the largest CO clusters vs time for $a=0$ and $a=1$. While spanning clusters were easily found in the $a=0$ case (around $45 \%$), only
around $0.05 \%$ were found to contain spanning clusters in the $a=1$ case.

\begin{figure}[h]
\includegraphics[width=0.8\textwidth]{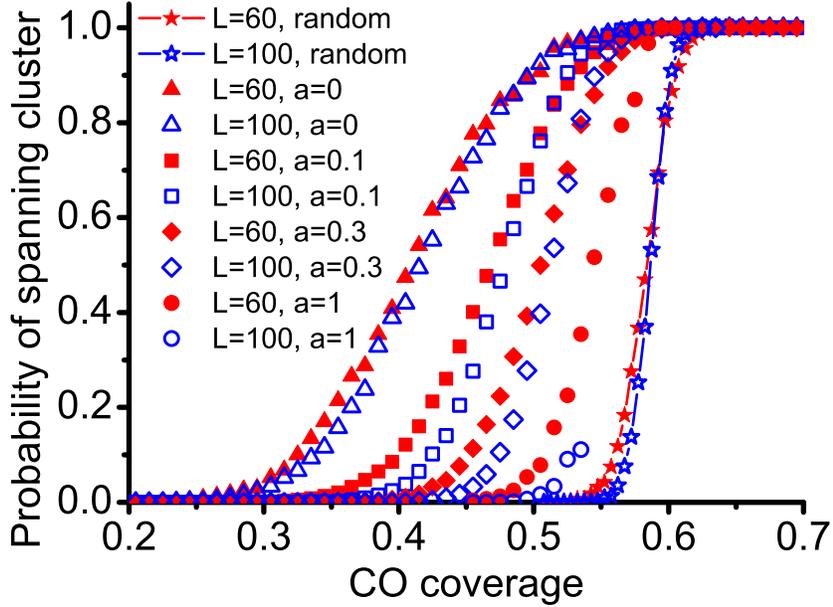}
\caption{(Color online) Dependence on the CO coverage of the
probability of finding a spanning cluster for $a=0$, $a=0.1$, $a=0.3$, and $a=1$ for
$60\times 60$ and $100\times 100$ lattices near their
critical points (i.e., at the $60/40$ and $100/60$ cumulant crossings, respectively), using a bin width of $0.01$.
Data were taken every $200$ MCSS for $a=0$, and every
$20$ MCSS for $a>0$.
$10^{8}$ MCSS were performed for $a=0$ and $10^{7}$ MCSS for $a>0$.
For $a=1$ and $L=60$, $59532$ snapshots were found to have CO coverage between $0.45$ and $0.61$, and among those $3714$ spanning clusters (6.24\%)
were found. For $a=1$ and $L=100$ only $17258$ snapshots were found to have CO coverage between $0.45$ and $0.61$, and among them there were only $23$ spanning clusters (0.13\%).
For $a=1$ we did not obtain useful data for CO coverages above $0.55$.
This effect is due to the narrowing with increasing $L$ of the critical CO-coverage distribution about its average at
approximately $0.33$. See Fig.~\protect\ref{fig_CO_coverage_distribution} and further discussion in the text.
Results for random percolation on $60\times 60$ and  $100\times 100$ lattices
are shown for comparison. For $a=0.1$ and 0.3 the data for the two system sizes display a clear
crossing. This suggests that the system in the mean-field case
develops a sharp percolation threshold that appears
to approach the random percolation threshold with increasing $a$.
}
\label{fig_percolation_critical}
\end{figure}

The large probability of spanning clusters for $a=0$ can of course be easily explained by the large number of
configurations with high CO coverages in this case (see Fig.~\ref{fig_cluster_size_gyration}(a)).
To get a meaningful picture, we must therefore compare
critical clusters in the $a=0$ and $a>0$ cases at the {\it same\/} CO concentration.
This is done in
Fig.~\ref{fig_percolation_critical}, which shows the probability of finding spanning clusters at the critical point vs
the CO coverage for the $a=0$ and three $a>0$ cases. This was obtained by sorting the snapshot configurations
according to their CO coverage in bins of width $0.01$ and plotting the relative number of spanning clusters in each bin.
The most striking feature of the figure is that percolation is rarer for configurations with a given CO coverage
at a mean-field critical point, than at the Ising critical point ($a=0$), an effect that becomes more pronounced with increasing $a$.

\begin{figure}[h]
\includegraphics[width=0.8\textwidth]{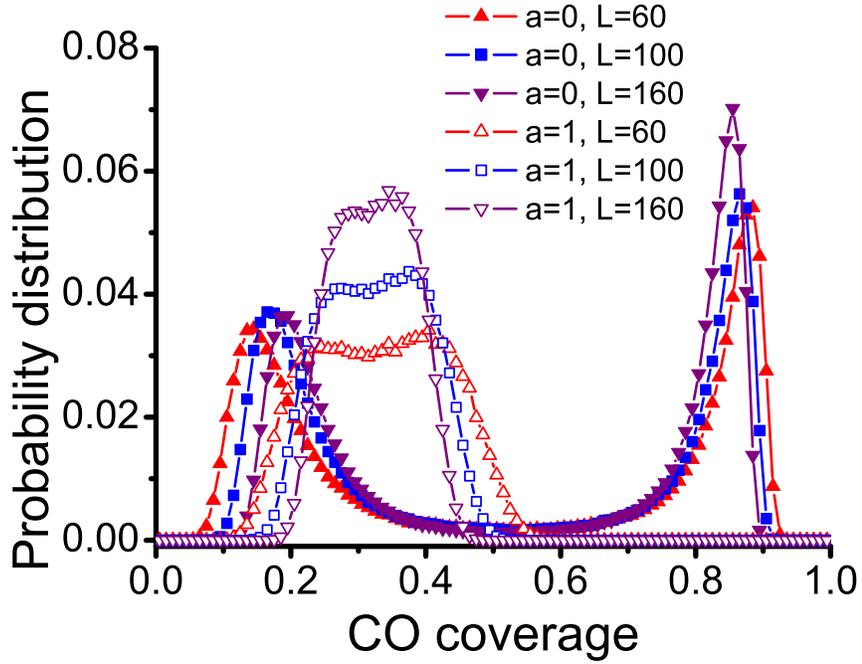}
\caption{(Color online) Critical probability distributions for the CO coverage plotted with a bin width of $0.01$.
The same sets of raw data were used for $L=60$, $100$ and $160$ at the critical points as in Fig.~\ref{fig_critical_CO_coverage}.
Data were taken every $200$ MCSS for $a=0$ and every $20$ MCSS for $a=1$. The mean-field case ($a=1$) shows unimodal distributions that narrow as
$L$ increases. In contrast, the Ising case ($a=0$) shows bimodal distributions with the two peaks shifting
slowly toward a central point as $L$ increases.
See further discussion in the text.
}\label{fig_CO_coverage_distribution}
\end{figure}

Results are shown in Fig.~\ref{fig_percolation_critical} for two system sizes, $L=60$ and $100$. The finite-size effects are
seen to be quite modest in the Ising case ($a=0$).
The CO coverage distribution for $a=1$ is a unimodal distribution (Fig.~\ref{fig_CO_coverage_distribution}) of mean CO coverage ($\langle\theta_{\rm{CO}}\rangle$) near
$0.33$ and with the average deviation from the mean CO coverage ($\langle|\theta_{{\rm CO},L} - \langle\theta_{{\rm CO},L}\rangle|\rangle$) expected to
decrease with increasing $L$ (see details in the next paragraph).
Very long simulations are therefore needed to
obtain reasonable statistics for CO coverages above $0.5$. As a result, we obtained results for CO coverages up to $0.61$
for $L=60$ in a run of $10^7$ MCSS, but only up to $0.55$ for $L=100$ using the same run length.
For $a=0.1$ and 0.3 the data for the two system sizes display a
clear crossing, as is also the case for random percolation \cite{NEWM01}. We interpret this as a sign that in the
mean-field case the system  develops a sharp percolation threshold that appears to approach the random
percolation threshold with increasing $a$.

The order-parameter distribution functions shown in Fig.~\ref{fig_CO_coverage_distribution} deserve some
further discussion. In the mean-field case ($a=1$), the distribution quickly approaches a unimodal form with an average near 0.33 as $L$ increases. Its width is expected to decrease with $L$ as $L^{-\beta / \nu}$ with
the mean-field critical exponents $\beta = 1/2$ and $\nu = 1$. Numerically we obtained   $\beta / \nu =
 0.609 \pm 0.009, 0.557 \pm 0.006, 0.553 \pm 0.004, 0.555 \pm 0.008$,
and $0.549 \pm 0.006 $ for $a=0.1,0.3,0.5,0.7$, and $1.0$, respectively.
We consider this consistent with the exact value of 0.5 for the mean-field universality class.
In contrast, the Ising case ($a=0$) shows bimodal distributions with the two peaks shifting
slowly toward a central point as $L$ increases.
The narrowing is expected to go as $L^{-\beta / \nu}$ with the Ising critical exponents $\beta = 1/8$
and $\nu = 1$.
Numerically we obtained $\beta / \nu = 0.0977 \pm 0.0007$ for $a=0$ using $L=60,100$, and $160$.
We consider this consistent with the exact value of 0.125 for the Ising universality class.
At the critical point, the two peaks should have equal weight of $50\%$ each.
Numerically we find that $48.7\%, 49.1\% $ and $44.0\%$ of the data points
have a CO coverage of less than $0.5$ for $L=60,100$, and $160$, respectively. These results are close to the expected
value of $50\%$. $L=160$ has a relatively larger deviation compared to $L=60$ and $100$ even though according to Fig.~\ref{fig_critical_intercept_a0_3graph}, the critical point for $L=160$ should be more accurately
determined than that for $L=60$. The reason is that the width of the critical region in the direction
perpendicular to the coexistence line (i.e., approximately in the $y$ direction) shrinks with $L$ as
$L^{-\beta \delta / \nu} = L^{-15/8}$ \cite{PRIV84,ROBB07}.
As a result, even a small deviation from the critical point can have a large
deleterious effect on the symmetry of the order-parameter distribution. This can be seen in the data point for
$a=0, L=160$ in Fig.~\ref{fig_critical_CO_coverage}, and it is even more pronounced for $L=240$ (not shown).

We suggest that a qualitative explanation for the differences between the finite-size effects
in Fig.~\ref{fig_percolation_critical} for the Ising and
mean-field cases can be found by considering the form of the spanning probability function
for random percolation on a square lattice of linear size $L$ \cite{NEWM01},
\begin{equation}
R_L(p) = \exp \left[ - c L (p_c - p)^{\nu_p} \right] \;.
\label{eq-rp}
\end{equation}
Here, $p$ is the site occupation probability, $p_c$ is the random percolation threshold ($\approx 0.593$ \cite{NEWM01}),
and $\nu_p$ is the critical exponent for the connectance length of the percolation problem
 ($= 4/3$ \cite{NEWM01}). Ignoring the effect of
correlations on the percolation threshold, we approximately map our correlated percolation problem onto random
percolation by replacing the system size $L$ by the effective size $\hat L = L/ \xi$, where $\xi$ is the critical
order-parameter correlation
length of the interacting model (not to be confused with the percolation connectance length).
For the Ising universality class, $\xi \sim L$ at criticality, indicating that the spanning probability for CO
coverages  below the (modified) percolation threshold should be (approximately) independent of $L$.
In contrast, the correlation length in the mean-field universality class approaches a constant value as $L$ increases
\cite{PhysRevB.84.054433,NAKA12}.
Consequently we expect that the $L$ dependence of Eq.~(\ref{eq-rp}) should also qualitatively describe the
behavior for $a=1$.
The rarity of large clusters is a well-known
feature of mean-field critical points in equilibrium models \cite{PhysRevB.77.014105,PhysRevB.84.054433,NAKA12}. These observations therefore further
strengthen our conclusion that any nonzero long-range reactivity induces mean-field behavior in this nonequilibrium system.
In Sec.~\ref{sec:correlation_measurement} below we confirm that the correlation lengths in the
models studied here indeed obey the $L$ dependence postulated in this paragraph on the basis of
the known behaviors in the corresponding equilibrium models.

\subsection{Correlation function and correlation-length measurements}\label{sec:correlation_measurement}
\begin{figure}[h]
\includegraphics[width=0.8\textwidth]{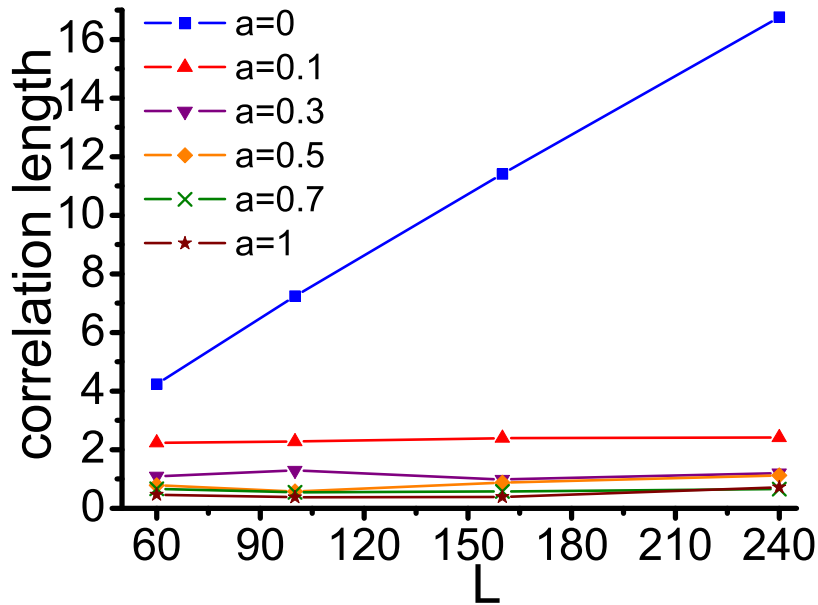}
\caption{(Color online) Correlation length for $a=0,0.1,0.3,0.5,0.7$, and $1$
at  their corresponding critical points (at the $60/40$, $100/60$ and $160/100$ cumulant crossings, respectively),
shown vs system size $L$. $10^{8}$ MCSS were used for $a=0$ and  $10^{7}$ MCSS for $a > 0$.
Without long-range reactivity ($a=0$), the correlation length increases linearly with $L$,
whereas in the presence of long-range reactivity ($a>0$), it is roughly independent of $L$ and decreases with increasing $a$.
}\label{fig_correlation_length}
\end{figure}

In order to verify the correlation-length scaling relations postulated in Sec.~\ref{sec:cluster_size_measurement}
above, we define the  CO disconnected correlation function as
\cite{PhysRevB.84.054433}
\begin{equation}
c(r) = \langle \sigma_{i} \sigma_{j} \rangle \;,
\label{eq-cf}
\end{equation}
where $\sigma_{i}$ is $1$ if site $i$ is occupied by CO and is $0$ otherwise, $r$ is the distance between site $i$ and site $j$, and the spatial
average is taken along the horizontal and vertical directions.
The critical correlation length is estimated by integration as
\begin{equation}\label{correlation_length}
 \xi(L) = \frac{\int^{L/2}_{0} [\langle c(r)\rangle - \langle c(L/2)\rangle] r dr  }{\int^{L/2}_{0} [\langle c(r)\rangle - \langle c(L/2)\rangle] dr } \;.
\end{equation}
As shown in Fig.~\ref{fig_correlation_length}, $\xi \sim L$ at the $a=0$ critical point, while it remains at
approximately $L$-independent values for $a>0$. These results are consistent with Ising critical behavior in
the former case and mean-field criticality in the latter.

\section{Conclusion}\label{sec:conclusion}
We employed large-scale Monte Carlo simulations using the crossing of fourth-order
cumulants to study the critical properties of the ZGB model with desorption, with and without long-range reactivity.
We obtained improved estimates for the critical point and the corresponding cumulant for the original ZBG model with CO desorption
($k_{c} = 0.0371 \pm 0.0002$, $y_{c} = 0.54052 \pm 0.00009$, $ u_{c,\infty} = 0.624 \pm 0.003$),
through the crossing of
cumulants up to a system size of $240 \times 240$ and run times of $4\times 10^{8}$ MCSS.
With the definition of the desorption rate used by Brosilow and Ziff \cite{PhysRevA.46.4534},
$P = k/(1-k)$, our result corresponds to
$P_c = 0.0385 \pm 0.0002$, close to the result obtained by those authors.

By adding long-range reactivity to the model, we find that the critical point of this nonequilibrium system changes from the
two-dimensional Ising universality class to the mean-field universality class. This change occurs even if the
long-range reactivity is quite weak. Our conclusion is supported by the fixed-point values of
fourth-order cumulants, as well as by the finite-size scaling behavior of the critical correlation length and by estimates of
the critical exponent ratio $\beta/ \nu$.
Moreover, while spanning clusters are easily observed near the critical point in the case without long-range reactivity, spanning clusters are seldom found near the critical point in the case with strong long-range reactivity.
This is so even when the cases are compared at the same value of the CO coverage.
The results of adding long-range reactivity to this nonequilibrium model are thus fully consistent with what
has previously been observed for weak long-range interactions in equilibrium Ising ferromagnets, providing
an example of the intriguing equivalence of critical phenomena in some equilibrium and nonequilibrium systems.

\section*{ACKNOWLEDGMENTS}\label{sec:acknowledgement}
We thank Gregory Brown, Tjipto Juwono, Yuhui Zhang, Alexander Gurfinkel, Gloria M. Buend{\'\i}a, Mark A. Novotry, and Yan Xu for discussions and help, and R.~M.\ Ziff for helpful correspondence and comments on the manuscript.
This work was supported in part by NSF Grant No. DMR-$1104829$.



\begin{thebibliography}{27}
\bibitem{Endnote1}
In some cases we find that parameters estimated from extrapolations of cumulant crossings based on the larger and the smaller
of the two system sizes involved in a crossing differ by more than their statistical uncertainties. This is an indication that even our large systems may not be fully in the asymptotic scaling region. In these cases (marked by an asterisk in Table~\ref{table_critical_pt}), we give the result based on the larger system sizes, but with the statistical error replaced by the difference between the two estimates.
\end{thebibliography}

\end{document}